\documentclass{amsart}
\usepackage{amssymb,amsmath,latexsym}
\usepackage[varg]{pxfonts}

\usepackage{graphicx}
\usepackage{subcaption}
\newtheorem{theorem}{Theorem}[section]

\newtheorem{corollary}[theorem]{Corollary}

\newtheorem{definition}[theorem]{Definition}
\newtheorem{example}[theorem]{Example}

\newtheorem{lemma}[theorem]{Lemma}

\usepackage{amssymb}
\usepackage{mathtools}
\usepackage{amsmath}

\newcommand{\R}{{\mathbb{R}}}

\newcommand{\N}{{\mathbb{N}}}

\title[DFA for continuous real variable functions]{Detrended Fluctuation Analysis for Continuous Real Variable Functions}
\author{Luis A.~Gil-Maqueda} 
\address{}
\email{luisgil.1105@gmail.com }
\author{Benjam\'in A.~Itz\'a-Ortiz}
\email{itza@uaeh.edu.mx}
\address{\'Area Acad\'emica de Matemáticas y Física\\ Universidad Autónoma del Estado de Hidalgo\\ Ctra.\ Pachuca-Tulancingo Km.\ 4.5\\ Pachuca, Hidalgo, Mexico 42184}

\begin{document}   

\begin{abstract}
Detrended Fluctuation Analysis (DFA) is a widely used technique for estimating scaling exponents and long-range correlations in time series data. In this work, we formulate DFA in a continuous framework (which we call Continuous DFA or CDFA for short) and provide a rigorous analytical description of its behavior for real-valued continuous functions. We derive an explicit integral representation of the fluctuation function and establish its regularity properties. We show that the fluctuation grows linearly at sufficiently small scales for all signals, so that the associated scaling exponent converges universally to one.

Although finite scaling ranges may exhibit apparent power law fits with different exponents, these reflect only local approximations rather than the true asymptotic behavior. Our results show that CDFA primarily measures first-order regularity rather than correlation structure. Consequently, while the classical discrete DFA distinguishes rough stochastic processes, CDFA loses discriminating power for continuous deterministic signals. These findings provide a clear theoretical explanation of both the capabilities and the intrinsic limitations of DFA.
\end{abstract}
 \maketitle     
    
   
   \section*{Introduction} 
\indent 

Time series analysis provides useful tools for understanding the dynamics of data collected from various research areas such as business, economics, medicine, and volcanology, among others \cite{ser1, ser3, ser2}. A fundamental question is whether the data exhibit temporal autocorrelation, meaning statistical dependence between observations at different times. In most processes these correlations decay rapidly, leading to a finite correlation length. A special case of particular interest is long-range dependence (or long memory) \cite{Ber}, where correlations decay slowly, typically according to a power law, so that distant observations remain significantly correlated across many scales. One way to quantify this behavior is through the Hurst exponent \cite{Hurst, Wynn}. In this context, 
Detrended Fluctuation Analysis (DFA) is widely used to estimate scaling behavior by measuring how detrended fluctuations of the integrated series grow with window size \cite{Peng}. DFA constructs the cumulative profile of the mean‑subtracted series, removes local polynomial trends on different scales, and computes the root‑mean‑square fluctuation $\mathcal F(n)$ as a function of scale $n$. A power law relation 
\[ \mathcal{F}(s) \propto s^{\alpha} \] indicates scale‑free correlations, and the scaling exponent  $\alpha$
 is related to the Hurst exponent. Thus, DFA provides a way to detect long‑range dependence even in nonstationary data, though not all signals yield a truly self‑similar profile. This method is widely used today, for example, in \cite{Berna, dfa1, dfa2, ser3,Peng, Itz, Peng2}, just to mention a few. On the other hand, it has also been criticized for presenting problems in mitigating nonstationarity \cite{BrySpr} or for not actually providing new information beyond spectral analysis \cite{Heneghan2000}, to name a few.

Since roughly speaking, DFA is a modified root mean square analysis, it is plausible to adapt its discrete framework to a continuous setting. A natural ensuing question, which we answer in the affirmative in this paper, is whether a continuous version of DFA also exhibits a power law behavior. In this work, we propose a DFA approach applied to continuous real-valued functions, which we call Continuous DFA (CDFA). 

The general idea of this work is to provide mathematical rigor to the proposed model from an analytical perspective, ranging from the fractal characteristics of the model~\cite{Hut, Hut2} to the derivation of a power law. In this setting, we do not work with signals exhibiting a random nature, but rather with continuous functions. 

As in the classical DFA for time series, the proposed CDFA  will consist of two steps. In the first step we associate an integrable function with an integrated one.  Recall that in the classical DFA, the cumulative sum  (i.e., the integrated series) of certain stochastic processes, such as fractional Gaussian noise, yields a self-similar process (fractional Brownian motion). In our  formulation of CDFA, we prove that the integrated function associated with any integrable function admits a decomposition as the sum of two self-similar fractal functions, where the notion of fractal function is taken in the sense of Hutchinson \cite{Hut, Hut2}.
In the second step, we restrict our attention to continuous real-valued functions and propose a function $\mathcal{F}$ that measures the global mean behavior of the integrated process obtained previously in step one. The function $\mathcal{F}$ plays a role analogous to the detrended fluctuation function in classical DFA. As in classical DFA, we prove that $\mathcal{F}$ behaves approximately as a power law $c\,r^\alpha$ for small values of $r$. Unlike classical DFA, we show that for continuous functions $\mathcal{F}$ follows a power law with exponent $1$ in the asymptotic sense; that is, $ \lim_{r \to 0^+} \frac{\mathcal{F}(r)}{r} = C \in (0,\infty)$. This reveals the presence of highly correlated values, which coincides with the fact that $x(s)$ is a deterministic object. With this, we note that when we consider the continuous model, it loses the capability to detect non-regular behavior as in the discrete case. At first sight, the fact that CDFA yields the same small-scale
exponent for a broad class of functions may seem to limit its
discriminatory power. However, this behavior should not be viewed as a
deficiency of the analysis but rather as a structural property of the
method itself. The results above show that any DFA procedure based on
local polynomial detrending necessarily reduces, in the continuous
limit, to a measurement of first-order smoothness. Consequently, the
collapse of the scaling exponent to one is universal and does not depend
on the particular signal considered. 

Thus, the purpose of the present work is not to construct a more powerful practical variant of DFA, but rather to provide a rigorous theoretical understanding of what DFA actually measures and to determine its intrinsic limitations when extended to continuous deterministic signals. We prove that, for any continuous deterministic function, the fluctuation function grows linearly at sufficiently small scales, so that the CDFA exponent necessarily converges to one. Consequently, CDFA no longer discriminates between different continuous deterministic signals. This clarifies why DFA is effective for rough or stochastic processes while CDFA loses discriminating
power in the continuous-time limit. Our approach follows the same motivation outlined by Veneziano \cite{veneziano} for testing multifractal analysis with analytic functions; however, in the case of DFA, the limitation arises precisely from the transition to continuous time.

We divide this work into two sections. In  Section~1, we review the concept of self-similar fractal functions and provide some examples of such functions. In Section~2 we present the main results of the paper,  namely, we introduce CDFA and prove that it approximates, in an asymptotic sense, a power law with an exponent equal to one.

The authors extend their sincere gratitude to the anonymous referee for the meticulous review of the manuscript. The referee not only recognized the value and relevance of the work, but also provided valuable comments and suggestions that helped refine various aspects of the paper and significantly improved its presentation.

\section{Self-similar Fractal Function}
In this section, we introduce the notion of self-similar fractal functions and present several examples. In classical DFA, the integrated series of certain stochastic models, such as fractional Gaussian noise,  yields fractional Brownian motion, which is self‑similar. For this reason, when extending the method to continuous functions, it is natural to consider an analogous notion of self-similarity for real-valued functions. We adopt the definition introduced by Hutchinson \cite{Hut, Hut2}, which provides a convenient functional counterpart to the theory of self-similar sets. This formulation facilitates the construction of examples and preserves several geometric properties of fractals, such as translation invariance.

\begin{definition} Let $N\geq 2$. 
A scaling law $\mathcal{S}$ is defined to be an $N-$tuple $(S_1,\dots S_N)$ of Lipschitz maps $S_i:\R^n\rightarrow \R^n$. We denote the Lipschitz constants by $\mathrm{Lip}(S_i)$.
\end{definition}

\begin{definition}
Let $I=I_1\sqcup \cdots\sqcup I_N$ be a partition of an interval $I$ into $N$ disjoint subintervals.  Given maps $g_i:I_i\rightarrow \R^n$ for $i=1,\dots , N$,  define the function $\bigsqcup^{N}_{i=1} g_i:I\rightarrow \R^n$ by
\begin{equation*}
    \left(\bigsqcup^{N}_{i=1} g_i\right)(x)=g_j(x),\quad x\in I_j.
\end{equation*}
\end{definition}

\begin{definition}\label{selfsim}
Let $f:I\rightarrow \R^n$ be a function where $I\subset \R$ is a closed bounded interval. Let $I=I_1\sqcup \cdots\sqcup I_N$ be a partition of $I$ into disjoint subintervals and let $\phi_i:I\rightarrow \overline{I_i}$ be an onto and increasing Lipschitz map for each $i\in \{1,\dots N\}$.
Given a scaling law $\mathcal{S}=(S_1,\dots ,S_N)$, define $\mathcal{S}f:I\rightarrow \R^n$ by
\begin{equation*}
    \mathcal{S}f= \bigsqcup_{i=1}^{N}S_i\circ f\circ \phi_i^{-1}.
\end{equation*}
We say that $f$ satisfies the scaling law $\mathcal{S}$, or that $f$ is a \textbf{self-similar fractal function} if
\begin{equation}\label{fracfuncio}
    f=\mathcal{S}f.
\end{equation}
\end{definition}
Note that in Definition~\ref{selfsim} the use of Lipschitz maps may be seen as  the analogous of verifying scaling properties in $f$, and by the disjoint union we can interpret that we are joining the pieces after the scaling, so this emulates the behavior of known fractals. Consider the following important example. 
\begin{example}\label{rec}
The function $f:[a,b]\rightarrow \R$ given by $f(x)=cx$, where $c\in \R$ is a constant, is a self-similar fractal function.
\begin{proof}
Consider the following disjoint partition of the interval $[a,b]=\left[a,\frac{a+b}{2}\right] \sqcup \left(\frac{a+b}{2},b\right]$ and let
\begin{equation*}
     \phi_1:[a,b]\rightarrow \left[a,\frac{a+b}{2}\right],\quad  \phi_1(x)=\frac{x}{2}+\frac{a}{2}
\end{equation*}
\begin{equation*}
     \phi_2:[a,b]\rightarrow \left[\frac{a+b}{2},b\right],\quad  \phi_2(x)=\frac{x}{2}+\frac{b}{2}.
\end{equation*}
It is  easy to see that $\phi_1$ and $\phi_2$ are  Lipschitz maps because are differentiable, and  derivative is positive.

Consider the scaling law $(S_1,S_2)=\left (\frac{x}{2}+\frac{a}{2c},\frac{x}{2}+\frac{b}{2c}\right)$, then a straightforward computation shows that
 \begin{equation*}
      Sf=\left(\bigsqcup^{2}_{i=2} S_i\circ f\circ \phi^{-1}_i\right)=cx.
 \end{equation*}
 and hence   $(\ref{fracfuncio})$ follows. Then $f$ is a self-similar fractal function.
\end{proof}
\end{example}
 The following lemma is relevant to establish  the translations invariance of self-similar fractal function.
\begin{lemma}\label{translema}
Let $I$ be a bounded and closed interval in $\R$ and let  $S:I\rightarrow \R$  be a Lipschitz map. Given a constant $c\in \R$, consider the set $Y=\{x+c : x\in I\}$. If we define $\Bar{S}:Y\rightarrow \R$ in $Y$ by $\Bar{S}(x+c)=S(x)+c$, then $\Bar{S}$ is a Lipschitz map.
\end{lemma}
\begin{proof}
Let $y_1,y_2\in Y$ i.e $y_1=x_1+c$ y $y_2=x_2+c$ then
\begin{align*}
|\Bar{S}(y_1)-\Bar{S}(y_2)|
=|S(x_2)-S(x_1)|\leq \mathrm{Lip}(S)|x_1-x_2|=\mathrm{Lip}(S)|y_1-y_2|.
\end{align*} 
Then, $\Bar{S}$ is a Lipschitz map, as wanted.
\end{proof}
\begin{theorem}\label{trasl}
Let $I$ be a closed bounded interval, and let $c\in \R$. If $f:I\rightarrow \R$ is a self-similar fractal function, then  $f+c$ is also a self-similar fractal function.
\begin{proof}
By hypothesis $f$ admits the following  representation
\begin{equation*}
f=\bigsqcup^{N}_{i=1}S\circ f \circ \phi^{-1}_i,
\end{equation*}
for some scaling law $(S_1,\dots ,S_N)$ and some family of Lipschitz maps $\phi_i$.

For every $i\in \{1,2,\dots ,N\}$ consider the family of sets  $Y_i=\{f \circ \phi^{-1}_i(x)+c\colon x\in I_i\}$, and let us define a function $\overline{S_i}$ in these sets, by $\Bar{S}_i(f \circ \phi^{-1}_i(x)+c)
=S_i(f \circ \phi^{-1}_i(x))+c$, which is Lipschitz by Lemma \ref{translema} for every $i\in \{1,2\dots ,N\}$. Then:
\begin{align*}
\bigsqcup^{N}_{i=1}\overline{S}\circ (f+c) \circ \phi^{-1}_i(x)
&=\bigsqcup^{N}_{i=1}\overline{S}(f \circ \phi^{-1}_i+c)\\
&=\bigsqcup^{N}_{i=1} \left(S(f \circ \phi^{-1}_i)+c\right)\\
&=\left(\bigsqcup^{N}_{i=1} S(f \circ \phi^{-1}_i)\right)+c\\
&=f+c.
\end{align*}
Hence, the function $f+c$ is a self-similar fractal function, with scaling law $(\Bar{S_1},\Bar{S_2},\dots,\Bar{S_N})$.
\end{proof}
\end{theorem}

\section{DFA and CDFA models}

The classical DFA can be viewed as a procedure that transforms a time series into a fluctuation function whose dependence on scale often exhibits approximate power law behavior. It consists of two main steps. In this section we present continuous analogues of these steps, thereby defining the Continuous DFA model (CDFA), and we provide a rigorous analysis of its properties and scaling behavior.


\subsection{First step: integrated function.}

The first step of the classical DFA is the following, consider a time series of size $M\in \N$, $x(i), i=1,\dots, M$, then define the integrated time series given by $y(i)=\sum_{j=1}^i\thinspace (x(j)-\overline{x})$, where $\overline{x}=\frac{1}{M}\sum_{i=1}^{M} x(i)$ is the average value of the time series. This new time series is a self-similar process.
\begin{definition}\label{IP}
Let $M$ be a positive real number. Suppose that $x:[0,M] \rightarrow \R$ is an integrable function. We define the \textbf{integrated function} of $x(t)$ to be the function $y\colon [0,M]\to \R$ given by the formula:
\begin{equation}\label{inte}
     y(t)=\int_{0}^{t} (x(s)-\Bar{x} )\ ds,
\end{equation}
where $\displaystyle \Bar{x} = \frac{1}{M} \int_{0}^{M} x(s) ds$.
\end{definition}
Notice that to a time series $a(i)$ of size $M\in \N$ corresponds a simple function $x:[0,M]\rightarrow \R$ given by $x(t)=\sum_{i=1}^M a(i)\chi_{[i-1,i]}(t)$. The averages of $a(i)$ and its corresponding simple function $x(t)$ coincide, that is, $\overline{a}=\overline{x}$. Furthermore, if $b(i)$ is the integrated time series of $a(i)$ and $y(t)$ is the integrated function of the corresponding simple function $x(t)$ of $a(i)$, then $y(t)=\sum_{i=1}^M b(i)\chi_{[i-1,i]}(t)$. Hence, the definition of an integrated function generalizes the notion of an integrated series.

In classical DFA, the integrated series of certain stochastic models (for example, fractional Gaussian noise) exhibits self-similar behavior. This motivates the question of whether, in the continuous setting, the integrated function can be represented as a self-similar fractal function in the sense of Definition~\ref{selfsim}.  Although such an exact representation does not generally hold, the next theorem shows that the integrated function can be decomposed as the sum of two functions of this type.

\begin{theorem}\label{fra}
Let $x:[0,M]\rightarrow \R$ be an integrable function with $M>0$. 
\begin{enumerate}
    \item If there exists $\delta>0$ such that $x(s)\geq \delta$ for all $s$, then $z\colon[0,M]\to\R$ defined by
    \begin{equation*}
        z(t)= \int_0^t x(s) \thinspace ds,
    \end{equation*}
    is a self-similar fractal function.
    
    \item The integrated function of $x(t)$ is the sum of two self-similar fractal functions.
\end{enumerate}
\end{theorem}
\begin{proof}
 To prove part (1), consider the disjoint partition of  $[0,M]=[0,\frac{M}{2}]\cup(\frac{M}{2},M]$ and the increasing Lipschitz maps $\phi_1$ and $\phi_2$ given by:
\begin{equation*}
    \phi_1:[0,M]\rightarrow \left[0,\frac{M}{2}\right],\quad \phi_1(t)=\frac{t}{2},
\end{equation*}
\begin{equation*}
    \phi_2:[0,M]\rightarrow \left[\frac{M}{2},M\right],\quad \phi_2(t)=\frac{t}{2}+\frac{M}{2}.
\end{equation*}
with inverse maps:
\begin{equation*}
    \phi^{-1}_1(t)=2t. 
\end{equation*}
\begin{equation*}
    \phi^{-1}_2(t)=2t-M.
\end{equation*}
Define the map $S_1$ on the image of  $z\circ \phi^{-1}_1$ by:
\begin{equation*}
    S_1(z(\phi^{-1}_1(t)))=S_1\left(\int_{0}^{2t} x(s)\thinspace  ds\right):=\int_{0}^{t} x(s)\thinspace ds,
\end{equation*}
 define the map $S_2$ on  the image of  $z\circ \phi_2^{-1}$ by:\\
\begin{equation*}
    S_2(z(\phi^{-1}_2(t)))=S_2\left(\int_{0}^{2t-M} x(s)\thinspace  ds\right):=\int_{0}^{t} x(s)\thinspace ds.
\end{equation*}
If we show that 
$S_1$ and $S_2$ are Lipschitz maps on the images of  $z(\phi^{-1}_1(t))$ and $z(\phi^{-1}_2(t))$ respectively, then by Kirzbraun Theorem \cite{fed}, we can extend these maps to Lipschitz maps on $\R$.
 
By continuity of $x(s)$ in the compact set $[0,M]$, there exist  $K$ and $k$ given by:
 \begin{equation*}
      K=\sup\left\{x(s)\colon s \in \left[0,\frac{M}{2}\right]\right\}
 \end{equation*}
 and
 \begin{equation*}
      k=\inf\left\{x(2s)\colon s \in \left[0,\frac{M}{2}\right]\right\}.
 \end{equation*}

 For $t, \tau\in [0,\frac{M}{2}]$ with $t<\tau$ we have the following:
 \begin{equation}\label{de1}
     \int_{t}^{\tau} x(s)\thinspace ds\leq \int_{t}^{\tau} K  ds=(\tau-t)K,
 \end{equation}
 and
 \begin{equation}\label{de12}
     k(\tau-t)=\int_{t}^{\tau} k ds\leq \int_{t}^{\tau} x(2s)\thinspace ds,
 \end{equation}
 
\noindent note that by hypothesis $x(s)\geq \delta > 0$ so we have $K>0$ y $k>0$. Then we may choose $R > 0$ such that: $K\leq 2Rk$. Hence, using $(\ref{de1})$ and $(\ref{de12})$ we have that
 \begin{equation}\label{de3}
     \int_{t}^{\tau} x(s)\thinspace ds\leq 2R\int_{t}^{\tau} x(2s)\thinspace ds.
 \end{equation}
 And using a variable change $\overline{s}=\frac{s}{2}$, to the integral on the right in $(\ref{de3})$ we obtain
 \begin{equation*}
     \int_{t}^{\tau} x(s)\thinspace ds\leq R\int_{2t}^{2\tau} x(s)\thinspace  ds.
 \end{equation*}
Since by hypothesis $x(s)\geq \delta > 0$, the integrals in the last inequality are non-negative, so we obtain the following. 
 \begin{equation}\label{deslipc}
     \left|\int_{t}^{\tau} x(s)\thinspace ds\right|\leq R \left|\int_{2t}^{2\tau} x(s)\thinspace ds\right|.
 \end{equation}
Then, if we add a zero to the integral on the left hand side of $(\ref{deslipc})$ we obtain
 \begin{equation*}
     \left|\int_{t}^{\tau} x(s)\thinspace ds\right|=\left|\int_{0}^{t} x(s)\thinspace ds -\left(\int_{0}^{t} x(s)\thinspace ds+\int_{t}^{\tau} x(s)\thinspace ds\right)\right|
 \end{equation*}
 \begin{equation*}
     =\left|\int_{0}^{t} x(s)\thinspace ds -\int_{0}^{\tau} x(s)\thinspace ds\right|,
 \end{equation*}
 and if we add a zero to the integral on right hand side of $(\ref{deslipc})$
 \begin{equation*}
     \left|\int_{2t}^{2\tau} x(s)\thinspace ds\right|=\left|\int_{0}^{2t} x(s)\thinspace ds-\left(\int_{0}^{2t} x(s)\thinspace ds+\int_{2t}^{2\tau} x(s)\thinspace  ds\right)\right|
 \end{equation*}
 \begin{equation*}
      =\left|\int_{0}^{2t} x(s)\thinspace ds-\int_{0}^{2\tau} x(s)\thinspace ds\right|.
 \end{equation*}
 With this $(\ref{deslipc})$ becomes:
 \begin{equation*}
     \left|\int_{0}^{t} x(s)\thinspace ds -\int_{0}^{\tau} x(s)\thinspace ds\right|\leq R \left|\int_{0}^{2t} x(s)\thinspace ds-\int_{0}^{2\tau} x(s)\thinspace ds\right|,
 \end{equation*}
 thus
 \begin{equation*}
     |S_1(z(\phi^{-1}_1(t)))-S_1(z(\phi^{-1}_1(\tau)))|\leq R|z(\phi^{-1}_1(t))-z(\phi^{-1}_1(\tau))|.
 \end{equation*}
This proves that $S_1$ is a Lipschitz map on the  image of $z\circ\phi^{-1}_1$.  The proof that $S_2$  is a  Lipschitz map on the image of $z\circ \phi^{-1}_2$ is analogous, so we omit it.

Then we have that
 \begin{equation*}
     \int_{0}^{t} x(s)\thinspace ds=\bigsqcup_{i=1}^{2}S_i\circ z\circ \phi^{-1}_i,
 \end{equation*}
i.e $z(t)$ is a self-similar fractal function.

Now to prove part (2), consider $\overline{x}=\frac{1}{M}\int_{0}^M x(s)ds$ and let $\delta>0$. Then  there exists $c>0$ such that $x(t)-\Bar{x}+c\geq \delta> 0$. Define $x_1(t)=x(t)-\Bar{x}+c$. From the previous part a) we have that,  $z_1(t)=\int \limits_{0}^{t} x_1(s)\thinspace\thinspace ds$ is a self-similar fractal function, in other words 
\begin{equation*}
    z_1(t)=\int_{0}^{t} x_1(s)\thinspace ds=\int_{0}^{t} (x(s)-\Bar{x}+c)\ ds = y(t)+ct,
\end{equation*}
where $y(t)$ is the integrated function of $x(s)$. On the other hand, by Example \ref{rec} the function $ct$ is a self-similar fractal function. Hence
\begin{equation*}
    y(t)=z_1(t)-ct.
\end{equation*}
  is  a sum of two self-similar fractal functions.  
\end{proof}
The following corollary establishes a sufficient condition for a function to be self-similar fractal.
\begin{corollary}\label{app2}
Let $x:[0,M]\rightarrow \R$ continuous and differentiable function, and suppose that there exists $\delta>0$ such that $\frac{dx}{dt}\geq \delta$, then $x(t)$ is a self-similar fractal function.
\end{corollary}
\begin{proof}
Since by the hypothesis $\frac{dx}{ds}\geq \delta >0$, it is possible to apply  Theorem \ref{fra}, then:
\begin{equation*}
    z(t)=\int_0^t \frac{d x}{ds} ds = x(t)-x(0),
\end{equation*}
is a self-similar fractal function, and then $x(t)$ is a self-similar fractal function by Theorem \ref{trasl}. 
\end{proof}
\subsection{Second step: Fluctuation function.}
The second step in the classical DFA model consists in removing the trend from the integrated time series $y(i)$ of a given time series $x(i)$, $i=1,\ldots , N$. To achieve this, we restrict the integrated time series $y(i)$ to subintervals of size $n$, with $1<n<N$. With the data in each window of size $n$, the line of least squares is calculated. 
The $y$-coordinate value of this line is denoted by $y_n(i)$. The process of removing the trend from the integrated time series $y(i)$ is
carried out by subtracting the value of $y_n(i)$ from each window. For each $n$, the characteristic length is obtained for the fluctuations of the integrated and trendless time series:
\begin{equation*}
    \mathcal{F}(n)=\sqrt{\frac{1}{N}\sum_{i=1}^N(y(i)-y_n(i))^2}.
\end{equation*}
From now on, we will require that the function $x$ be continuous and not just integrable as in the previous step. Since continuous functions on closed intervals are integrable,  we fulfill the conditions to define the integrated functions described in subsection \ref{IP}, and we also fulfill the hypothesis of  Theorem~\ref{fra}.

We now introduce a method to analyze the average behavior of the increments of the integrated function given in Definition~\ref{IP}, associated with a given continuous function $x\colon [0,M]\to\mathbb{R}$, where $M$ is a positive number. If one attempts to follow the classical DFA procedure, the trend would be removed by least-squares fitting on subintervals of fixed length. More precisely, the integrated function $y(t)$ would be restricted to windows of length $n$, with $0<n<M$, and on each window approximated by a linear (or, more generally, polynomial) function, whose value is then subtracted to obtain the detrended signal.
However, in the continuous setting this approach does not produce a power law scaling of the fluctuations, even when higher-order polynomial fits are used. For this reason, we adopt a different detrending strategy, based on an approach analogous to differentiation.

Let $0<m<M$. Restrict the integrated function $y(t)$ given in Definition~\ref{IP} to the interval $[m,M]$ and let $0<r<m$. Consider the difference
\begin{equation}\label{TR}
     y(t)-y(t-r) = \int_{0}^{t} (x(s)-\Bar{x} ) ds - \int_{0}^{t-r} (x(s)-\Bar{x} ) ds = \int_{t-r}^{t} (x(s)-\Bar{x} ) ds.
\end{equation}
The expression in \eqref{TR} may be regarded as the process of removing the trend of $y(t)$, taking into account the immediate past, rather than the tendency established by the windows. Then, as in the methodology introduced by Peng et al. \cite{Peng}, consider the square root of the average of the squared values obtained in (\ref{TR}) as a function of $r$.  This gives the detrended function $\mathcal{F}\colon [0,m]\to\mathbb R$ defined by
\begin{align}\label{detr}
    \mathcal{F}(r)&:=\sqrt{\frac{1}{(M-m)}\int _{m}^{M} \left(  y(t)-y(t-r) \right)^2  dt}.
\end{align}

We next show that when $x$ is continuous, then $\mathcal F$ is also continuous. 
\begin{lemma}\label{detrendcont}
Assume that $x\colon[0,M]\to\R$ is a continuous function. Let $0<m<M$ and consider the detrended  function $\mathcal{F}\colon[0,m]\to\R$ defined in \eqref{detr}. Then $\mathcal F^2$ is Lipschitz. In particular, $\mathcal F$ is  continuous.
\end{lemma}
\begin{proof}
Since $x$ is continuous on the compact interval $[0,M]$, it is bounded. 
Hence there exists $C>0$ such that
\[
|x(s)-\bar x| \le C
\quad \text{for all } s \in [0,M].
\]

For $r \in [0,m]$ and $t \in [m,M]$, define
\[
G(r,t)
:=
\int_{t-r}^{t} (x(s)-\bar x)\,ds.
\]
Then
\[
|G(r,t)|
\le
\int_{t-r}^{t} |x(s)-\bar x|\,ds
\le
Cr,
\]
and for any $r,r_0 \in [0,m]$,
\[
G(r,t)-G(r_0,t)
=
\int_{t-r}^{t-r_0} (x(s)-\bar x)\,ds,
\]
so that
\[
|G(r,t)-G(r_0,t)|
\le
C|r-r_0|.
\]
Using $|a^2-b^2| = |a-b||a+b|$ and the previous bounds,
\[
|G(r,t)^2 - G(r_0,t)^2|
\le
C|r-r_0| \cdot C(r+r_0)
\le
2C^2 m |r-r_0|.
\]
Integrating over $t$ gives
\[
|\mathcal F(r)^2 - \mathcal F(r_0)^2|
=
\frac{1}{M-m}
\left|
\int_m^M
\big(G(r,t)^2 - G(r_0,t)^2\big)\,dt
\right|
\le
2C^2 m |r-r_0|.
\]
Thus $\mathcal F^2$ is Lipschitz on $[0,m]$, hence continuous. Since the square root is continuous on $[0,\infty)$ and $\mathcal{F} \ge 0$, it follows that $F$ is continuous on $[0,m]$.
\end{proof}

We now focus on the asymptotic scaling behavior of the model $\mathcal F$. We say that $\mathcal{F}\neq 0$ behaves asymptotically as $r^\alpha$ if 
\begin{equation*}
\lim_{r\to 0^+}\frac{\mathcal{F}(r)}{r^\alpha}=C>0.
\end{equation*}
This means that, for small values of $r$, the function $\mathcal{F}(r)$ decays exactly at the rate $r^\alpha$. In this sense, the exponent $\alpha$ determines the order of vanishing of $\mathcal{F}$, while the constant $C$ quantifies the precise speed of this decay. This allows us to determine the behavior of $\mathcal F$ while taking into account the scale. 

If $x$ is continuous, in the next theorem we establish that the detrended fluctuation function satisfies
\[
\lim_{r\to 0^+} \mathcal F(r)=0,
\]
which simply reflects that a continuous signal becomes locally constant on sufficiently small windows, so that detrended fluctuations vanish. More importantly, we shall prove that
\[
\lim_{r\to 0^+} \frac{\mathcal F(r)}{r} = C < \infty,
\]
so that
\[
\mathcal F(r) = Cr + o(r),
\]
or equivalently ,
\[
\mathcal F(r)\to Cr \quad\text{as }r\to 0^+.
\]

Thus, the leading-order behavior of $\mathcal F$ near the origin is always linear. Consequently, any log--log regression of $\mathcal{F}(r)$ versus $r$ on sufficiently small scales necessarily yields an effective scaling exponent equal to one. This asymptotic statement should be distinguished from the purely topological fact that a continuous function vanishing at the origin can approximate $c r^\alpha$ in arbitrarily small intervals for any $\alpha>0$. Such approximations do not describe the leading-order scaling of $\mathcal F$, which is uniquely determined by the linear term above. Hence, for continuous inputs, CDFA measures local smoothness rather than genuine scale-invariant fluctuations. 
We are now ready to state and prove our main result.
\begin{theorem}\label{poteF}
    Let $M>0$ and let $x\colon [0,M]\rightarrow \R$ be a continuous function.  Let $0<m<M$ and consider the detrended function $\mathcal{F}:[0,m]\rightarrow \R$ given by \eqref{detr}.
Then $\mathcal{F}(r)$ is continuous and    
\begin{equation*}
   \lim_{r\to 0^+} \mathcal{F}(r)=0.
\end{equation*}
Moreover
\begin{equation}\label{asym}
    \lim_{r\to 0^+} \frac{\mathcal{F}(r)}{r} =\sqrt{\frac{1}{M-m}\int_{m}^{M}\left(x\left(t\right)-\overline{x}\right)^2\, dt}.
\end{equation}
In particular, $\lim_{r\to 0^+} \frac{\mathcal{F}(r)}{r}$ is a nonnegative constant number and equals 0 if and only if $x(t)=\Bar{x}$ for all $t\in[0,M]$.
\end{theorem}
\begin{proof}
By Lemma~\ref{detrendcont} the function $\mathcal F$ is right-continuous at $r=0$. Since $\mathcal F(0)=0$, we conclude that 
\begin{equation*}
    \lim_{r\to0^{+}}\mathcal{F}(r)=\mathcal{F}(0)=0.
\end{equation*}
To establish \eqref{asym}, it will suffice to prove the limit for an arbitrary sequence of positive numbers converging to zero. For this purpose, let $\{r_n\}_{n\geq 0}$ be a sequence of positive real numbers such that $r_n\to 0$ as $n\to\infty$. 
\newline
First, observe that by using \eqref{TR} we get
\[
\frac{\mathcal{F}(r)^2}{r^2}=\frac{1}{M-m}\int_{m}^{M}\biggl( \frac{1}{r}\int_{t-r}^{t}\left(x(s)-\overline{x}\right)\, ds\biggr)^2dt.
\]
Let 
\begin{equation*}
  f_n(t)=\frac{1}{r_n}\int_{t-r_n}^{t} \left(x(s)-\Bar{x}\right)ds.
\end{equation*}
Then, for a fixed $t\in[0,M]$
\begin{equation*}
    f_n(t)- \left( x(t)-\Bar{x}\right)=\frac{1}{r_n}\int_{t-r_n}^{t}\left(x(s)-x(t)\right)ds.
\end{equation*}
Since $x$ is continuous on a compact interval, it is uniformly continuous.
Let $\epsilon>0$. Then there exists $\delta>0$ such that $|s-t|<\delta$ implies $|x(s)-x(t)|<\epsilon$ for every $s$. Choose $N>0$ such that $r_n<\delta$ for every $n>N$. Then for $n>N$ and for every $s\in [t-r_n,t]$ it follows that $|x(s)-x(t)|<\epsilon.$ Hence
\begin{equation*}
   \left|\frac{1}{r_n}\int_{t-r_n}^{t}\left(x(s)-x(t)\right)ds\right|\leq \frac{1}{r_n}\int_{t-r_n}^{t}|x(s)-x(t)|ds\leq\frac{1}{r_n}\int_{t-r_n}^{t}\epsilon\, ds=\epsilon.
\end{equation*}
This proves 
\begin{equation*}
   \lim_{n\to \infty} \bigl(f_n(t)- \left( x(t)-\Bar{x}\right)\bigr)=\lim_{n\to\infty}\frac{1}{r_n}\int_{t-r_n}^{t}\left(x(s)-x(t)\right)ds=0.
\end{equation*}
Thus
\begin{equation*}
\lim_{n\to\infty}f_n(t)=x(t)-\Bar{x}.
\end{equation*}
and so
\begin{equation*}
f_n(t)^2\to\left(x(t)-\Bar{x}\right)^2  
\end{equation*}
pointwise on $[m,M].$

Using again that $x(s)$ is continuous on a compact interval we find that it is bounded. Therefore, there exists $C>0$ such that for every $s\in[0,M]$,
\begin{equation*}
    |x(s)-\Bar{x}|<C. 
\end{equation*}
We obtain
\begin{equation*}
    |f_n(t)|\leq\frac{1}{r_n}\int_{t-r_n}^{t}|x(s)-\Bar{x}|\,ds\leq\frac{1}{r_n}\int_{t-r_n}^{t}C\,ds=C.
\end{equation*}
It follows that the sequence of functions $\{f_n^2\}_{n\geq 0}$ converges pointwise to the integrable function $\left(x(t)-\Bar{x}\right)^2$ and is dominated by an
integrable (constant) function. By Lebesgue's dominated convergence theorem, we obtain
\begin{equation*}   \lim_{n\to\infty}\int_{m}^{M}f_n^2(t)\,dt=\int_{m}^{M}\lim_{n\to\infty}f_{n}^{2}(t)\, dt=\int_{m}^{M}\left(x(t)-\Bar{x}\right)^2\, dt
\end{equation*}
In summary,
\begin{align*}
   \biggl(\lim_{r\to 0^{+}}  \frac{\mathcal{F}(r)}{r}\biggr)^2&= \lim_{r\to 0^{+}}  \frac{\mathcal{F}(r)^2}{r^2} \\
   &=\frac{1}{M-m}\ \lim_{r\to 0^{+}}\int_{m}^{M}\biggl( \frac{1}{r}\int_{t-r}^{t}\left(x(s)-\overline{x}\right)\, ds\biggr)^2dt\\
   &=\frac{1}{M-m}\int_{m}^{M}\int_{m}^{M}\left(x(t)-\Bar{x}\right)^2\, dt.
\end{align*}
Hence
\begin{equation*}
    \lim_{r\to 0^+} \frac{\mathcal{F}(r)}{r} =\sqrt{\frac{1}{M-m}\int_{m}^{M}\left(x\left(t\right)-\overline{x}\right)^2\, dt},
\end{equation*} 
 as wanted.  
 \end{proof}

Consequently, at sufficiently small scales the fluctuation function is asymptotically linear. Therefore, the CDFA scaling exponent converges to one for any continuous signal. In particular, CDFA cannot discriminate between different continuous signals. This shows that the discriminating power of classical DFA relies essentially on stochastic roughness or discrete sampling rather than on continuous deterministic structure.

Now we present a brief numerical application of the continuous model to certain continuous functions not only to illustrate Theorem ~\ref{poteF} but also to reaffirm the structure limitation of DFA when considering CDFA as the extension to continuous signals.
\begin{figure}[ht]
    \centering
    \includegraphics[width=0.7\linewidth]{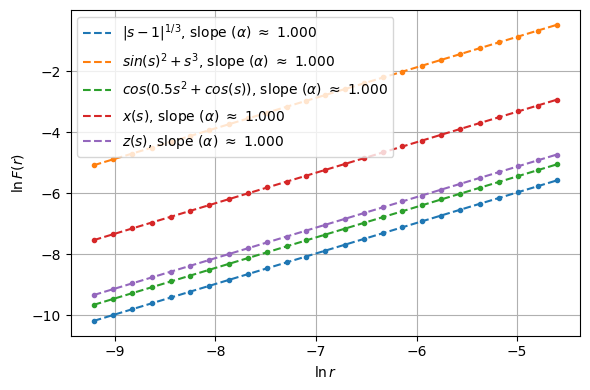}
    \caption{Fluctuation function $\mathcal{F}$ for several continuous test functions. The curves exhibit linear behavior at small scales, consistent with the asymptotic prediction of Theorem~\ref{poteF}.}
    \label{fig:Fclasica}
\end{figure}

\begin{figure}[htb]
    \centering
    \begin{subfigure}[b]{0.45\textwidth}
        \centering
        \includegraphics[width=\textwidth]{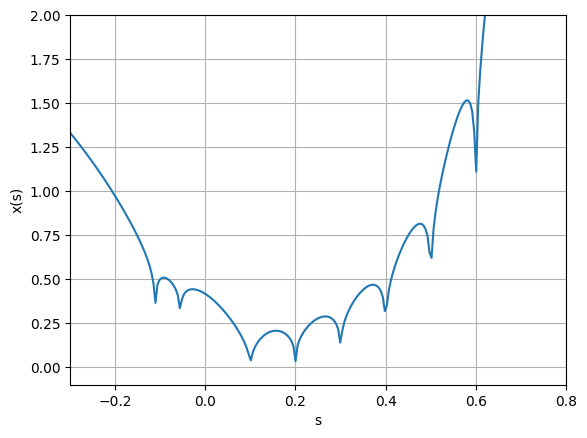}
        \caption{Plot of the function $x.$}
        \label{fig:a}
    \end{subfigure}
    \hfill
    \begin{subfigure}[b]{0.45\textwidth}
        \centering
        \includegraphics[width=\textwidth]{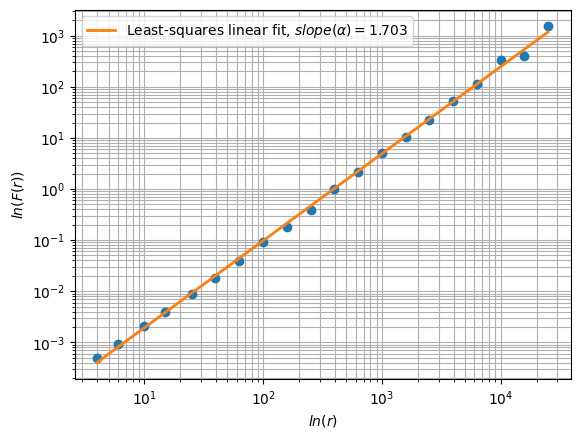}
        \caption{Log--log plots of classical $\mathcal{F}(r)$ applied to discretizations of $x$.}
        \label{fig:b}
    \end{subfigure}

    \medskip

    \begin{subfigure}[b]{0.45\textwidth}
        \centering
        \includegraphics[width=\textwidth]{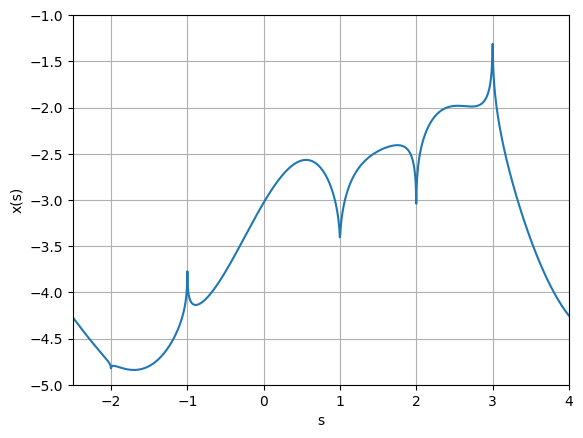}
        \caption{Plot of the function $z.$}
        \label{fig:c}
    \end{subfigure}
    \hfill
    \begin{subfigure}[b]{0.45\textwidth}
        \centering
        \includegraphics[width=\textwidth]{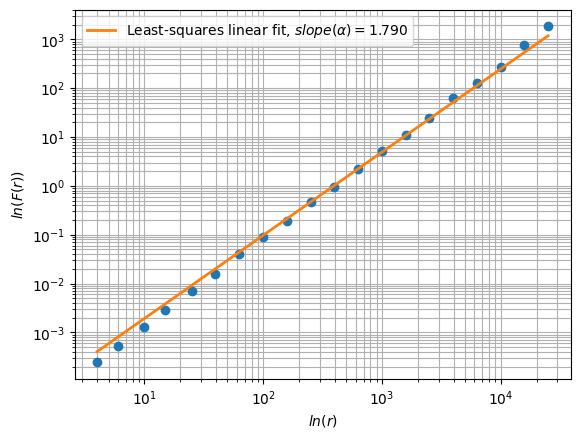}
        \caption{Log--log plots of classical $\mathcal{F}(r)$ applied to discretizations of $z$.}
        \label{fig:d}
    \end{subfigure}
    \caption{From continuous to discrete analysis. The function $\mathcal{F}$ denotes the detrended function defined in the classical DFA model giving exponents other than one.}
    \label{fig:4fig}
\end{figure}

In Figure ~\ref{fig:Fclasica}, we apply CDFA to five different continuous functions. The expressions for the first three are given in the figure, while the last two,  $x(s)$ and $z(s)$, are given  by 
\begin{equation}\label{multiplesfunc}
x(s)=
(s+2)^3
\left| s + \frac{1}{9} \right|^{1/8}
\left| s + \frac{1}{18} \right|^{1/9}
\left| s + \frac{1}{3} \right|^{1/7}
\prod_{i=1}^5
\left| s - \frac{i}{2} \right|^{1/(i+1)}
\end{equation}
and
\begin{equation}\label{multiplesfunc2}
    z(s)=\sin\bigl(2|x-1|^{1/2}|x-2|^{1/3}\bigr) - (3|x-1|)^{1/6}
    + \frac{1}{5}|x+2|^{1/7} - 2|x-2|^{1/4},
\end{equation}
respectively.
Note that although $|s-1|^{1/3}$, $x(s)$ and $z(s)$ have non-differentiable points, Theorem~\ref{poteF} establishes that the obtained exponent is $1$. This indicates that the model does not detect such non-differentiable behavior. It is known that if we extract a time series by considering a discrete finite set of values from the continuous function and apply classical DFA analysis to this time series, the resulting exponent is not necessarily one.
For example, let us consider finite discrete values $x(t_i)$ and $z(t_i)$, $i=1,\dots,N$, and apply the classical DFA. In Figure~\ref{fig:4fig}, we see that, in this case, the resulting exponents lie in the interval $(1.5,2)$. Indeed, this reflects the absence of Brownian noise and the smoothness of the signals, which is consistent with the behavior of $x$ and $z$. This illustrates that in the transition from continuous to discrete analysis, the classical DFA is able to distinguish the roughness of the signal.

\begin{figure}[ht]
    \centering
    \begin{subfigure}[b]{0.45\textwidth}
        \centering
        \includegraphics[width=\textwidth]{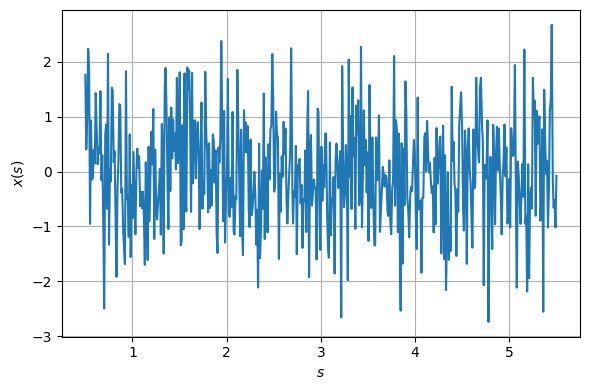}
        \caption{The function $l_1$ is the interpolation of data generated as white noise in Python.}
        \label{fig:e}
    \end{subfigure}
    \hfill
    \begin{subfigure}[b]{0.45\textwidth}
        \centering
        \includegraphics[width=\textwidth]{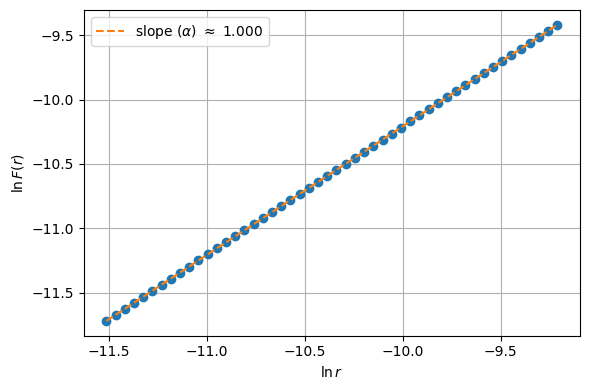}
        \caption{Log--log plots of $\mathcal{F}(r)$ applied to $l_1$.}
        \label{fig:f}
    \end{subfigure}
    \medskip

    \centering
    \begin{subfigure}[b]{0.45\textwidth}
        \centering
        \includegraphics[width=\textwidth]{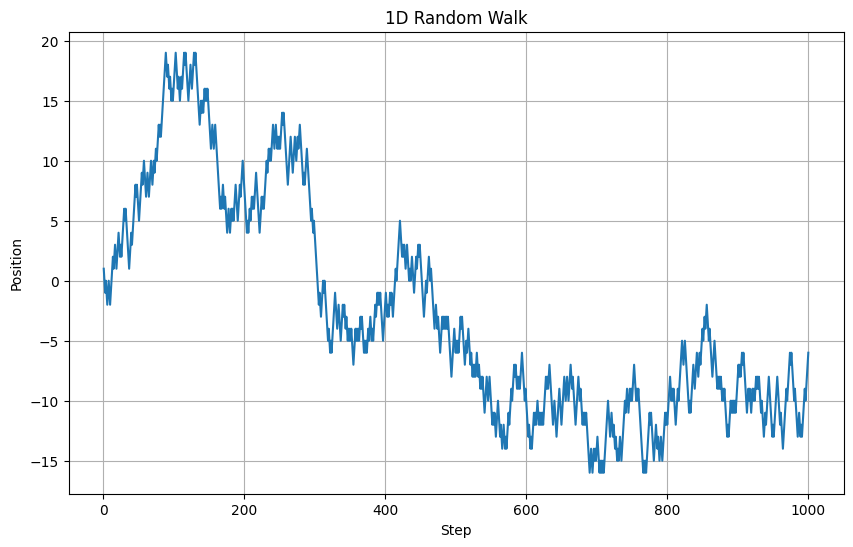}
        \caption{The function $l_2$ is the interpolation of data generated as a random walk ($p=1/2$)  in Python.}
        \label{fig:g}
    \end{subfigure}
    \hfill
    \begin{subfigure}[b]{0.45\textwidth}
        \centering
        \includegraphics[width=\textwidth]{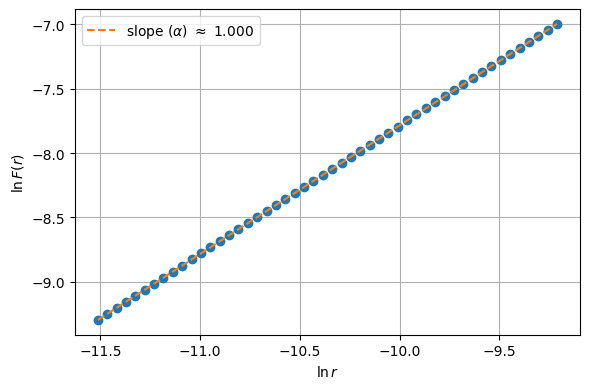}
        \caption{Log--log plots of $\mathcal{F}(r)$ applied to $l_2$.}
        \label{fig:h}
    \end{subfigure}
    \caption{From discrete to continuous  analysis. Illustration of how the discrete DFA procedure collapses in the continuous formulation (CDFA) introduced in this work.}
    \label{fig:dos_figuras}
\end{figure}

On the other hand, one may consider a given time series for which classical DFA  has a given exponent $\alpha$. Next, consider a continuous realization of the time series. The obtained function, when CDFA is applied, has exponent equal one. For example, consider data sets following a white noise $\{s_i\}_{i=1,\dots,500}$ and 
a random walk $\{t_j\}_{j=1,\dots,1000}$. After a linear interpolation we obtain continuous functions $l_1(s)$ and $l_2(t)$ such that $l_1(i)=s_i$ and $l_2(j)=t_j$. Then Theorem~\ref{poteF} establishes that the exponent will be equal to $1$. Numerically, we confirm this result in Figure~\ref{fig:dos_figuras}. This illustrates that the transition to CDFA leads to a collapse of the exponent that characterizes the behavior of the discrete
data, highlighting an intrinsic limitation of DFA when extended to continuous deterministic signals.
\section{Conclusions}
 The  CDFA model introduced in this work provides a rigorous analytic description of the method and clarifies what quantity it actually measures. While classical discrete DFA often exhibits apparent power law behavior with process-dependent exponents, we prove that in the continuous deterministic setting the fluctuation function is asymptotically linear at small scales, forcing the associated scaling exponent to converge universally to one. Consequently, CDFA is unable to discriminate between different continuous deterministic signals.

This result shows that the effectiveness of DFA relies essentially on stochastic roughness or discrete sampling rather than on smooth continuous structure. Our analysis therefore establishes a precise theoretical explanation of both the capabilities and the intrinsic limitations of DFA, and clarifies the behavior of the method when passing from discrete to continuous time.

\bibliography{Bibliography} 
\bibliographystyle{amsplain}





\end{document}